\documentclass[sigconf]{acmart}
\usepackage{enumitem}
\usepackage{subfigure}
\usepackage{threeparttable,multicol,multirow}
\usepackage{amsmath}
\usepackage{balance}

\copyrightyear{2022}
\acmYear{2022}
\setcopyright{acmcopyright}\acmConference[CIKM '22]{Proceedings of the 31st ACM International Conference on Information and Knowledge Management}{October 17--21, 2022}{Atlanta, GA, USA}
\acmBooktitle{Proceedings of the 31st ACM International Conference on Information and Knowledge Management (CIKM '22), October 17--21, 2022, Atlanta, GA, USA}
\acmPrice{15.00}
\acmDOI{10.1145/3511808.355708910.1145/3511808.3557089}
\acmISBN{978-1-4503-9236-5/22/10}

\begin{document}

\title{Temporal and Heterogeneous Graph Neural Network for Financial Time Series Prediction}

\author{Sheng Xiang}
\affiliation{%
  \institution{AAII}
  \institution{University of Technology Sydney}
  \city{Sydney}
  \state{Australia}
}
\email{sheng.xiang@student.uts.edu.au}

\author{Dawei Cheng}\authornote{Corresponding Author.}
\affiliation{%
  \institution{Department of Computer Science}
  \institution{Tongji University}
  \city{Shanghai}
  \state{China}
}
\email{dcheng@tongji.edu.cn}

\author{Chencheng Shang}
\affiliation{%
  \institution{HSBC Business School}
  \institution{Peking University}
  \city{Beijing}
  \state{China}
}
\email{pkuscc@stu.pku.edu.cn}

\author{Ying Zhang}
\affiliation{%
  \institution{AAII}
  \institution{University of Technology Sydney}
  \city{Sydney}
  \state{Australia}
}
\email{ying.zhang@uts.edu.au}

\author{Yuqi Liang}
\affiliation{%
  \institution{Seek Data Group}
  \institution{Emoney Inc.}
  \city{Shanghai}
  \state{China}
}
\email{roly.liang@seek-data.com}


\renewcommand{\shortauthors}{Sheng Xiang et al.}

\begin{abstract}
The price movement prediction of stock market has been a classical yet challenging problem, with the attention of both economists and computer scientists. In recent years, graph neural network has significantly improved the prediction performance by employing deep learning on company relations. However, existing relation graphs are usually constructed by handcraft human labeling or nature language processing, which are suffering from heavy resource requirement and low accuracy. Besides, they cannot effectively response to the dynamic changes in relation graphs. Therefore, in this paper, we propose a temporal and heterogeneous graph neural network-based (THGNN) approach to learn the dynamic relations among price movements in financial time series. In particular, we first generate the company relation graph for each trading day according to their historic price. Then we leverage a transformer encoder to encode the price movement information into temporal representations. Afterward, we propose a heterogeneous graph attention network to jointly optimize the embeddings of the financial time series data by transformer encoder and infer the probability of target movements. Finally, we conduct extensive experiments on the stock market in the United States and China. The results demonstrate the effectiveness and superior performance of our proposed methods compared with state-of-the-art baselines. Moreover, we also deploy the proposed THGNN in a real-world quantitative algorithm trading system, the accumulated portfolio return obtained by our method significantly outperforms other baselines.

\end{abstract}



\maketitle

\section{Introduction}
Stock market is a financial ecosystem that involves transactions between companies and investors, with a global market capitalization of more than \$83.5 trillion as of 2020~\cite{word2021bank}.
The Efficient Market Hypothesis~\cite{Merello2019EnsembleAO} points out that the stock price represents all the available information, but the stock price is volatile in nature, resulting in difficulty on predicting its movement~\cite{Adam2008StockMV}.
In recent years, deep learning has been widely used in predicting the price movement of stocks~\cite{Li2020Elstm}. Researchers also explore to improve the prediction performance by incorporating more sources as model inputs, including more technical indicators~\cite{Merello2019EnsembleAO}, factors \cite{Feng2019EnhancingSM}, financial status~\cite{Ballings2015EvaluatingMC}, online news~\cite{liang2020f}, social media posts \cite{websciWangLHL19}, etc.

Traditional learning methods treat the time series as independent and identically distributed to each other, which is not coincident to the real situation in the financial market \cite{Li2019IndividualizedIF}. For example, two stocks in the same sector may have a higher correlation than those in different fields \cite{Chen2018IncorporatingCR}. Therefore, recent studies employ knowledge graphs to represent the internal relations among entities \cite{Cheng2020KnowledgeGE,zhu2022leveraging} and leverage graph learning for the price movement prediction~\cite{cikmLiu0SZ18}. These works have shown the effectiveness by integrating stock's relations in the prediction models \cite{Cheng2021ModelingTM}. Thus, graph-based methods could benefit from learning meaningful representations of inputs, resulting in better prediction accuracy \cite{cheng2019dynamic,Sawhney2020DeepAL}.

However, generating relation graphs of entities is very challenging because it is fluctuate, noisy, amphibolous and dynamically changed \cite{Li2020ModelingTS}. For example, financial knowledge graphs mainly include the supply chain, primary business and investment relations of entities \cite{Feng2019TemporalRR}, which is labeled by domain experts or extracted from unstructured texts. But different experts have different knowledge background which may lead to different relation graphs.
Besides, the current nature language processing (NLP) techniques are still facing significant shortcomings in high accuracy relation extraction \cite{emnlpDingZLD14}. In other words, the relations may be misled by either unilateral text news or inaccurate extracting models.
In addition, these relations may dynamically change in time series. For example, the main business of a company would change according to the market demands, and the supply chain graph would be upgraded because of the technique evolution \cite{houston2016financial}. Existing graph learning price prediction methods are inevitably suboptimal in learning these fluctuate and dynamical situations.

To address the above challenges, we propose a novel temporal and heterogeneous graph neural network-based method for financial time series prediction. 
Specifically, we directly model the relations of price time series of entities based on historical data and represent them in a temporal and heterogeneous graph, i.e., company relational graph. After obtaining the company relational graph, we leverage sequential transformers encode the historical prices and graph neural networks to encode internal relations of each company. Specifically, we update each company's representations by aggregating information from their neighbors in company relational graphs in two steps. The first step is a time-aware graph attention mechanism. The second is a heterogeneous graph attention mechanism.
We thoroughly evaluate our approach on both the S\&P 500\footnote{https://www.spglobal.com/spdji/en/indices/equity/sp-500}  and CSI 300\footnote{http://www.cffex.com.cn/en\_new/CSI300IndexOptions.html}  dataset in the United States and China's stock markets. The experimental results show that our method significantly outperforms state-of-the-art baselines. In order to keep the sententious of our model, we conduct ablation studies to prove the effectiveness and essential of the each component of our method, including transformer encoder, time-aware graph attention, heterogeneous graph attention. Finally, we deploy our model in real-world quantitative algorithm trading platform, hosted in EMoney Inc.\footnote{http://www.emoney.cn/}, a leading financial service provider in China. The cumulative returns of portfolios contributed by our approach is significantly better than existing models in financial industry. We will release the dataset as well as the source codes of the proposed techniques along with the paper. In conclusion, our principle contributions are summarized as follows:
\begin{itemize}
   \item We propose a graph learning framework to effectively model the internal relations among entities for financial time series prediction, which fits the dynamical market status and is concordant with the ground-truth price movements.
   \item We design a temporal and heterogeneous graph neural network model to learn the dynamic relationships by two-stage attention mechanisms. The proposed model is concise and effective in joint and automatically learning from historical price sequence and internal relations.
   \item Our proposed THGNN is simple and can be easily implemented in the industry-level system. Extensive experiments on both the Unite States and China stock markets demonstrate the superior performance of our proposed methods. We also extensively evaluated its effectiveness by real-world trading platform.
\end{itemize}


\section{Related Works}

\subsection{Financial Time Series Learning}
It is widely known that price movements of stocks are affected by various aspects in financial market \cite{Feng2019EnhancingSM}. In previous studies, a common strategy is to manually construct various factors as feature inputs~\cite{CHENG2022108218,finrl_meta_2021}. For example, Michel et. al, \cite{Ballings2015EvaluatingMC} integrate market signals with stock fundamental and technical indicators to make decisions. Li et. al, \cite{Li2020ModelingTS} establish a link between news articles and the related entities for stock price movement forecasting. A large number of existing methods employ recurrent neural network and its variants, such as LSTM~\cite{Hochreiter1997LongSM} and GRU~\cite{Cho2014LearningPR}, to learn the sequential latent features of historical information and employ them for downstream prediction task \cite{Jin2019StockCP}.

In these works, the market signals processing of each stock is carried out independently. However, this inevitably ignores the internal relationship among stocks and would lead to suboptimal performance. Some works \cite{ganeshapillai2013learning} leverage the correlation information as model inputs, but cannot automatically capture the dynamic changes of relations. In this article, we model the relationship between stocks as dynamic company relation graphs and joint learn the graph relation and historical sequence feature automatically for future price movement prediction.

\subsection{Graph Learning for Stock Prediction}
Researchers have shown that the price movement of stocks is not only related to its own historical prices, but also connect to its linked stocks \cite{Cheng2021ModelingTM}. The link relation includes suppliers and customers, shareholders and investor, etc. Existing works normally employ knowledge graphs to store and represent these relations \cite{liu2021item,CHENG2022108218}. Recently, graph neural network (GNN) \cite{wu2020comprehensive} is proposed to effectively learn on graph-structured data, which has shown its superior performance in various domains, including fraud detection \cite{cheng2020delinquent,cheng2020graph}, computer vision \cite{tu2020image,Tu2020LearningFW}, etc. Researchers also introduce the advanced GNN-based approaches in the stock price prediction task. For example, Chen et. al, \cite{Chen2018IncorporatingCR} models the supply chain relationships of entities into knowledge graphs and uses graph convolution networks to predict stock movements.
Ramit et. al, \cite{sawhney2020deep} leverage attentional graph neural network on the connections constructed by social media texts and company correlations. Cheng et.al, \cite{CHENG2022108218} leverage multi-modality graph neural network on the connections constructed by historical price series, media news, and associated events.

However, the graph constructed by these methods are limited by constant, predefined corporate relationships, which is powered by handcraft editing or nature language processing techniques,  suffering from heavy resources labeling and low extracting accuracy \cite{ji2021survey}. But the actual corporate diagram evolves frequently over time. Besides, the company relation graph is also heterogeneous, which means there are multiple relation types among entities. Therefore, existing methods cannot exploit the full information from real-life company relation graphs. In this paper, we construct the relation graph dynamically and automatically based on their ground-truth historical price sequences and then propose a novel temporal and heterogeneous graph neural network methods to jointly learn their sequential and relational features for more accurate stock price prediction. We demonstrate the effectiveness of our methods by extensive experiments and real-world trading platform.

\section{The Proposed Method}
In this section, we introduce the framework of our proposed temporal and heterogeneous graph neural network and their each component in detail. Our model takes the historical price sequence as inputs and infer the probability of stock movements as outputs. We represent the relation of stocks in a dynamic heterogeneous graph with two types of edges. We then jointly encode the historical and relation features by transformers and heterogeneous graph attention network. We report the problem definition first and each module of our method in turn.

\renewcommand\arraystretch{1.0}
\begin{table}\vspace{-0pt}
  \caption{The summary of symbols}\label{tb:symbols}\vspace{-5pt}
  \centering
  \begin{tabular}{|c|l|}
     \hline
     \textbf{Symbol} & \textbf{Definition} \\ \hline \hline
     $\mathbf{X}$ & the historical price of listed companies\\ \hline
     $\mathbf{\hat{Y}}$ & the probability of price movements\\ \hline
     $n$ & the total number of nodes \\ \hline
     $m$ & the total number of edges \\ \hline
     $r$ & the number of relationships in graph $\Tilde{G}$ \\ \hline
     $T$ & the number of trading days in $\Tilde{G}$\\ \hline
     $\Tilde{G}=(\Tilde{V},\{\Tilde{E}_{r_1},\Tilde{E}_{r_2},...\})$ & the temporal and heterogeneous graph \\ \hline
     $\Tilde{V}=\{v_1^{t_{v_1}},...,v_n^{t_{v_n}}\}$ & the set of temporal nodes\\ \hline
     $\Tilde{E}_r=\{e_1^{t_{e_1}},...,e_m^{t_{e_m}}\}_r$ & the set of temporal edges of relation $r$ \\ \hline
     $\mathcal{N}(\cdot)$ & the neighborhood function\\ \hline
     $d$ & the number of dimension\\ \hline
  \end{tabular}
  \vspace{-5pt}
\end{table}

\begin{figure*}[tb!]\vspace{-10pt}
   \centering
   \includegraphics[width=0.99\linewidth]{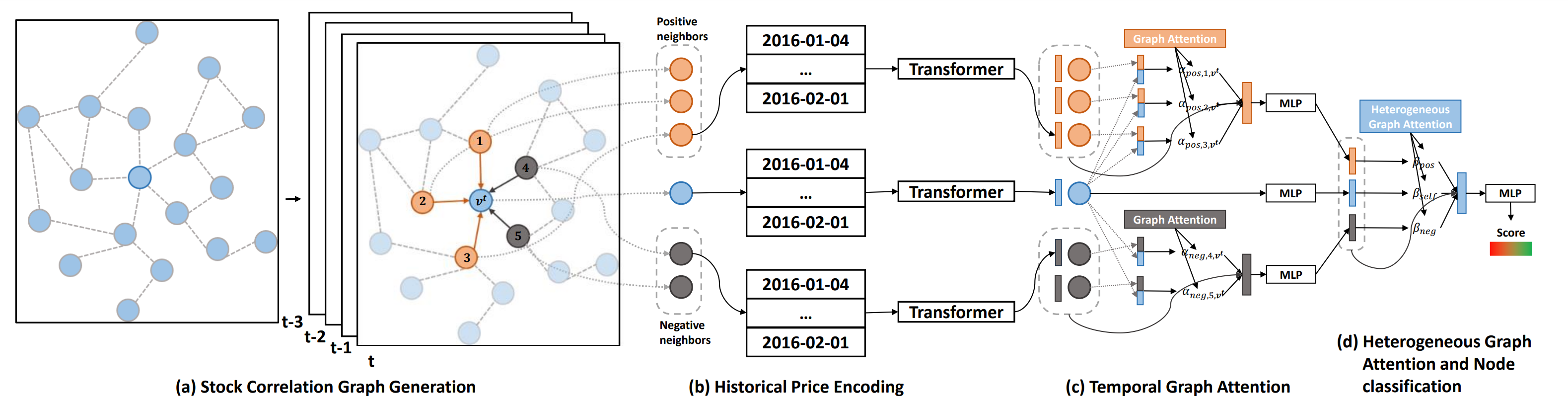}\vspace{-10pt}
   \caption{
   The proposed Temporal and Heterogeneous Graph Neural Networks architecture for stock movement predictions. The first part is the generation of a stock correlation graph, which builds dynamic relations for stocks in the market every trading day. The second part is the historical price encoding, which selects a temporal node $v^t$ and its neighbor nodes to encode the historical price information. Transformer encoders share their parameters. The third part is the graph attention layer, which adaptively calculates the importance of the neighbors and aggregates the information according to the neighbors' importance. The fourth part is the heterogeneous graph attention layer, which adaptively calculates the importance and aggregates information from different types of neighbors. Then, we leverage a multi-layer perception to give the prediction of each stock's future movement.
   }
   \label{fig:framework}\vspace{-10pt}
\end{figure*}

\subsection{Problem Definition}
Different from traditional graph-based methods that construct static and homogeneous graphs by handcraft labeling or nature language processing techniques to infer stock movements, our model represents the company relation graph as a collection of temporal and heterogeneous graphs, which are automatically generated by historical price sequences. In graphs, the node denotes each equity and edge represents their relations in sequential. Temporal graphs are composed of timestamped edges and timestamped nodes~\cite{Zhou2020ADG,Paranjape2017MotifsIT}.
Each node might be associated with multiple relationships and multiple timestamped edges on different trading days. There are multiple types of edge $E=\{E_1,...,E_r\}$ in company relation graphs. And the occurences of node $V=\{V^{t_1},...,V^{T}\}$ and edge $E=\{E^{t_1},...,E^{T}\}$ are different on different trading days. Table~\ref{tb:symbols} summarizes the symbols introduced in this paper.

\begin{definition}
    \label{Temporal and Relational Occurrence}
    \textbf{Temporal and Relational Occurrence.} \textit{In a temporal company relation graph, an edge $e$ is associated with a series of temporal occurrences $e=\{e^{t_1},e^{t_2},...\}$, which indicate the occurrences of edge $e$ at trading days $\{t_1, t_2,...\}$ in the company relation graph. Each type of relational occurrence is associated with a series of temporal occurrences with $E=\{E_{r_1},E_{r_2},...\}$, which indicate the occurrences of edge $e$ in different relationships $\{r_1,r_2,...\}$. Same as temporal occurrences of edge $e$, a node $v$ is associated with a set of temporal occurrences with $v=\{v^{t_1},v^{t_2},...\}$.}
\end{definition}

\begin{definition}
    \label{Temporal and Heterogeneous Graph}
    \textbf{Temporal and Heterogeneous Graph.} \textit{A temporal and heterogeneous company relation graph $\Tilde{G}=(\Tilde{V},\Tilde{E})$ is formed by a set of temporal nodes $\Tilde{V}=\{v_1^{t_{v_1}},...,v_n^{t_{v_n}}\}$ and a series of sets of temporal edges $\Tilde{E}=\{\Tilde{E}_1, ...,\Tilde{E}_r\}$, where  $\Tilde{E}_r=\{e_1^{t_{e_1}},...,e_m^{t_{e_m}}\}_r$ denotes the edges of relation $r$, and $e_i^{t_{e_i}}=(u_{e_i},v_{e_i})^{t_{e_i}}$ denotes the temporal edge.}
\end{definition}

In existing works~\cite{Chen2018IncorporatingCR,Feng2019TemporalRR,Cheng2021ModelingTM}, the graph neighborhood $\mathcal{N}(v)$ of node $v$ is defined as static or homogeneous. Here, we generalize the definition of graph neighborhood to the temporal and heterogeneous graph, which is set as follows:

\begin{definition}
    \label{Temporal and Heterogeneous Graph Neighborhood}
    \textbf{Temporal and Heterogeneous Graph Neighborhood.} \textit{Given a temporal node $v$, the neighborhood of $v$ is defined as $\mathcal{N}(v)=\{v_i|f_{sp}(v_i,v)\leq d_{\mathcal{N}},|t_v-t_{v_i}|\leq t_{\mathcal{N}}\}$}, where $f_{sp}(\cdot|\cdot)$ denotes the shortest path length between two nodes, $d_{\mathcal{N}}$ and $t_{\mathcal{N}}$ denote the hyper-parameters. As for heterogeneous graph, we define $\mathcal{N}_r(\cdot)$ as the neighborhood function of relation $r$.
\end{definition}

Finally, we formally define the stock movement prediction problem as follows:

\textbf{Input:} \textit{Historical price sequences of listed companies $\mathbf{X}=\{x_1,x_2,\cdots,x_n\}$, where each $x_i= \{x_{i,1},x_{i,2},\cdots,x_{i,t}\}$ denotes the historical price sequences. We then use the $\mathbf{X}$ to generate the temporal and heterogeneous company relation graph $\Tilde{G}$, with multiple types of temporal edges $\{\Tilde{E}_{r_1},\Tilde{E}_{r_2},...\}$, for downstream tasks }.

\textbf{Output:} \textit{The probability $\mathbf{\hat{Y}}$ of price movements of each equity.}

\subsection{Stock Correlation Graph Generation}
\label{sec:cormat}
In this subsection, we report the process of the temporal and heterogeneous graph construction.
As mentioned in previous studies~\cite{Chen2018IncorporatingCR,Kim2019HATSAH}, there may be multiple relationships between companies (such as suppliers and customers, shareholders and invested companies). Different from conventional knowledge graph-based relations that construct relation by human labeling or NLP techniques, generating relations directly based on market trend signals has proved to be effective~\cite{Li2020ModelingTS,Xiang2021GeneralGG} in practical, which does not require additional ambiguity domain knowledge or text news sources, and is easy to be implemented.

Therefore, in this paper, we obtain the correlation matrix by calculating the correlation coefficients between ground-truth stock historical market signals directly.
Then, the relationship between companies is determined according to the value of each element of the correlation matrix.
The relationship between companies may be positive (correlation $>$ threshold) or negative (correlation $<$ -threshold).
In order to reduce noise, we connect the edges whose absolute value is greater than the threshold, and the rest of the edges are considered not to be connected. So far, the edges $E$ of company relation graph are generated.
Therefore, we model the inter-company relation graph as a heterogeneous graph with two relationships, i.e., $G=(V,\{E_{r_1},E_{r_2},...\})$, with $r\in \{pos,neg\}$.

As the relationships between companies tend to be dynamic, we generate the company relation graph in temporal format as our model's inputs. In particular, within $T$ trading days, we generate temporal and heterogeneous company relation graphs with $T$ timestamps, which is formulated as $\Tilde{G}=(\Tilde{V},\{\Tilde{E}_{pos},\Tilde{E}_{neg}\})$. Finally, the generated graphs and original sequence inputs are fed to downstream learning task simultaneously.

\subsection{Historical Price Encoding}
The input stock movement feature of price sequences is defined as $\mathbf{X}^t\in \mathbb{R}^{n\times T\times d_{feat}}$ on trading day $t$, where $n$ denotes the number of stocks, $T$ means the number of trading days before $t$, and $d_{feat}$ denotes the dimension of historical price features. We first leverage a linear transformation and positional encoding (PE)~\cite{Vaswani2017AttentionIA,Ying2021DoTR} on trading features to obtain the input tensor $\mathbf{H}^t\in \mathbb{R}^{n\times T\times d_{in}}$, which is formulated as follows:
\begin{equation}
    \begin{split}
        \hat{\mathbf{H}}^t=&\mathbf{W}_{in}\mathbf{X}^t+\mathbf{b}_{in}\\
        \mathbf{H}^t=&\hat{\mathbf{H}}^t+\text{PE}\\
        \text{PE}(p,2i)=&sin(p/10000^{2i/d_{in}})\\
        \text{PE}(p,2i+1)=&cos(p/10000^{2i/d_{in}})\\
    \end{split}
\end{equation}
where $p\in\{1,2,..,T\}$ is the trading day position, $i$ is the dimension, $d_{in}$ denotes the dimension of input features, and $\mathbf{W}_{in}\in \mathbb{R}^{d_{feat}\times d_{in}}$ and $\mathbf{b}_{in}\in \mathbb{R}^{d_{in}}$ denote the learnable parameters. After linear transformation, we proposed to leverage multi-head attentional transformer to encode the input feature for each stock in each day. 
Then, the proposed encoder outputs $\mathbf{H}_{enc}^t\in\mathbb{R}^{n\times T\times d_{enc}}$ for downstream tasks, where $d_{enc}$ denotes the output dimension of the encoder. Mathematically, we formulate the historical feature encoder's output $\mathbf{H}_{enc}^t$ as follows:
\begin{equation}
    \begin{split}
        \mathbf{H}_{enc}^t=\text{Concat}(\text{EncHead}^t_1,…,\text{EncHead}^t_{h_{enc}})\mathbf{W}_o
    \end{split}
\end{equation}
where $\mathbf{W}_o\in \mathbb{R}^{h_{enc}d_{v}\times d_{enc}}$ denotes the output projection matrix, $h_{enc}$ denotes the number of heads in the encoder, $d_v$ denotes the output dimension of each head, Concat denotes a concatenation of the output of heads, and $\text{EncHead}^t_i\in \mathbb{R}^{n\times T\times d_v}$ denotes the output of encoder head with $\text{EncHead}^t_i = \text{Attention}(\mathbf{Q}^t_i, \mathbf{K}^t_i, \mathbf{V}^t_i)$, which is formulated as follows:
\begin{equation}
    \begin{split}
        \mathbf{Q}^t_i=\mathbf{H}^t\mathbf{W}_i^Q, \mathbf{K}^t_i=&\mathbf{H}^t\mathbf{W}_i^K, \mathbf{V}^t_i=\mathbf{H}^t\mathbf{W}_i^V\\
        \text{Attention}(\mathbf{Q},\mathbf{K},\mathbf{V})=&\text{softmax}(\frac{\mathbf{Q}\mathbf{K}^T}{\sqrt{d_{in}}})\mathbf{V}
    \end{split}
\end{equation}
where $\mathbf{W}_i^Q\in\mathbb{R}^{d_{in}\times d_{hidden}}$, $\mathbf{W}_i^K\in\mathbb{R}^{d_{in}\times d_{hidden}}$, $\mathbf{W}_i^V\in\mathbb{R}^{d_{in}\times d_{v}}$ denote the projection matrices, and $d_{hidden}$ denotes the dimension of hidden layer.

\subsection{Temporal Graph Attention Mechanism}
\label{sec:tga}
Given the historical sequence encoder output $\mathbf{H}^t_{enc}$ and temporal relation graph $\Tilde{G}$, we propose to employ graph attention mechanism on the sequential and heterogeneous inputs.
In particular, we flatten the embeddings of all nodes to $\mathbf{H}^t_{enc}\in \mathbb{R}^{n\times Td_{enc}}$ and leverage two-stage temporal attention mechanism to aggregate messages from graph structures and temporal sequences, which is illustrative reported in Figure~\ref{fig:framework} (c). The two-stage temporal graph attention layers could aggregate messages from both the postive and negative neighbors simultaneously. For each relationship $r\in\{pos,neg\}$, the message aggregating is formulated as follows:
\begin{equation}
    \begin{split}
        \mathbf{H}_r^t=&\text{Concat}(\text{TgaHead}_1,...,\text{TgaHead}_{h_{tga}})\textbf{W}_{o,r}\\
    \end{split}
\end{equation}
where $\mathbf{H}_r^t\in \mathbb{R}^{n\times d_{att}}$ denotes the output of the temporal graph attention layer on trading day $t$, and $\textbf{W}_{o,r}\in \mathbb{R}^{h_{tga}Td_{enc}\times d_{att}}$ denotes the output projection matrix, $h_{tga}$ denotes the number of heads, and each head of temporal graph attention layer $\text{TgaHead}_i\in\mathbb{R}^{n\times Td_{enc}}$ is formulated as follows:
\begin{equation}
    \begin{split}
        \text{TgaHead}_i=\sum_{v^t\in\Tilde{V}}\sigma(\sum_{u^t\in \mathcal{N}_{r}(v^t)}\alpha_{u^t,v^t}^i\mathbf{h}_{u^t})\\
    \end{split}
\end{equation}
where $\sigma$ denotes the activation function, $\mathbf{h}_{u^t}\in\mathbb{R}^{Td_{enc}}$ denotes the $u^t$-th row of historical price embedding $\mathbf{H}^t_{enc}$, and $\alpha_{u^t,v^t}^i$ denotes the importance of temporal edge $(u^t,v^t)$ in $i$-th head, which is formulated as follows:
\begin{equation}
    \begin{split}
        \alpha_{{u}^t,{v}^t}^i=\frac{\text{exp}(\text{LeakyReLU}(\mathbf{a}_{r,i}^T[\mathbf{h}_{u^t}||\mathbf{h}_{v^t}]))}{\sum_{{k}^t\in \mathcal{N}_{r}({v}^t)}\text{exp}(\text{LeakyReLU}(\mathbf{a}_{r,i}^T[\mathbf{h}_{k^t}||\mathbf{h}_{v^t}]))} \\
    \end{split}
\end{equation}
where $\mathbf{a}_{r,i}\in\mathbb{R}^{2Td_{enc}}$ denotes the weight vector of relation $r$ and $i$-th head.

\subsection{Heterogeneous Graph Attention Mechanism}
As shown in Figure~\ref{fig:framework} (d), we already have messages from different types of neighbors through the two-stage attention mechanism. Then, we propose the heterogeneous graph attention network  to learn from different relationships in relation graphs.

We define message sources as three types of embeddings, namely, messages from ourselves $\mathbf{H}^t_{self}$, positive neighbors $\mathbf{H}^t_{pos}$, and negative neighbors $\mathbf{H}^t_{neg}$, respectively. $\mathbf{H}^t_{self}\in\mathbb{R}^{n\times d_{att}}$ is derived from $\mathbf{H}^t_{enc}$ through a linear transformation with $\mathbf{H}^t_{self}=\mathbf{W}_{self}\mathbf{H}^t_{enc}+\mathbf{b}_{self}$, where $\mathbf{W}_{self}\in\mathbb{R}^{Td_{enc}\times d_{att}}$ and $\mathbf{b}_{self}\in\mathbb{R}^{d_{att}}$ denote the learnable parameters. $\mathbf{H}^t_{pos}$ and $\mathbf{H}^t_{neg}$ are derived from the graph attention mechanism in section~\ref{sec:tga}. Taking three groups of node embeddings as input, we can adaptively generate the importance of different relationships through attention mechanism. The weights of three relationships $(\beta_{self},\beta_{pos},\beta_{neg})$ can be shown as follows:
\begin{equation}
    \begin{split}
        (\beta_{self},\beta_{pos},\beta_{neg})=\text{HGA}(\mathbf{H}^t_{self},\mathbf{H}^t_{pos},\mathbf{H}^t_{neg})\\
    \end{split}
\end{equation}

We first use three Multi-Layer Perceptrons (MLP) to transform these three embeddings. Then we measure the importance of each embedding using a heterogeneous attention vector $\mathbf{q}$. Furthermore, we
average the importance of all node embeddings, which can be explained as the importance of each company relation. The importance of each company relation, denoted as $r\in\{self,pos,neg\}$, is shown as follows:
\begin{equation}
    \begin{split}
        w_r=\frac{1}{|\Tilde{V}|}\sum_{v^t\in\Tilde{V}}\mathbf{q}^T\text{tanh}(\mathbf{W}\mathbf{h}_{v^t,r}+\mathbf{b})
    \end{split}
\end{equation}
where $\mathbf{W}\in\mathbb{R}^{d_{att}\times d_q}$ and $\mathbf{b}\in\mathbb{R}^{d_q}$ are the parameters of MLP, $\mathbf{q}\in\mathbb{R}^{d_q}$ denotes the attention vector, and $\mathbf{h}_{v^t,r}$ denotes the $v_t$-th row of $\mathbf{H}^t_{r}$. Note that all above parameters are shared for all relationship of node embeddings. After obtaining the importance of each relationship, we calculate the contribution of each relationship and obtain the final embedding $\mathbf{Z}^t\in\mathbb{R}^{n\times d_{att}}$ as follows:
\begin{equation}
    \begin{split}
        \beta_r=&\frac{\text{exp}(w_r)}{\sum_{r\in \{self,pos,neg\}}\text{exp}(w_r)}\\
        \mathbf{Z}^t=&\sum_{r\in \{self,pos,neg\}}\beta_r\cdot\mathbf{H}^t_r
    \end{split}
\end{equation}
For a better understanding of the aggregating process of heterogeneous graph attention layer, we give a brief explanation in Figure~\ref{fig:framework} (d). Then we apply the final embedding to a semi-supervised node classification task.

\subsection{Optimization Objectives}
Here we give the implementation of objective functions.
We model the stock movement prediction task as a semi-supervised node classification problem. Specifically, we selected 200 stocks of which future movements are ranked in top-100 or bottom-100, and labeled the corresponding nodes as $1$ and $0$, respectively.
Then, we use one layer of MLP as the classifier to get the classification results of labeled nodes.
Furthermore, we use binary cross-entropy to calculate the objective function $L$, which is formulated as follows:
\begin{equation}
    \begin{split}
        \mathbf{\hat{Y}}=&\sigma(\mathbf{W}\mathbf{Z}^t_l+\mathbf{b})\\
        \mathcal{L}=&-\sum_{l\in \mathcal{Y}_t}[\mathbf{Y}^t_l\log(\mathbf{\hat{Y}}_l)+(1-\mathbf{Y}^t_l)\log(1-\mathbf{\hat{Y}}_l)]\\
    \end{split}
\end{equation}
where $\mathcal{Y}_t$ denotes the set of labeled nodes, $\mathbf{Y}^t_l$ and $\mathbf{Z}^t_l$ denote the label and embedding of the labeled node $l$, respectively, $\sigma$ denotes the Sigmoid activation function, and $\mathbf{W}$ and $\mathbf{b}$ are the parameters of MLP. With the guide of labeled data, we use Adam~\cite{Kingma2015AdamAM} optimizer to update the parameters of our proposed method. Please note that we use this objective function to jointly optimize the parameters of historical price encoder, temporal and heterogeneous graph neural network and node classifier.

\section{Experiments}
\label{sec:exprm}

In this section, we first introduce the datasets and experimental settings.
Then we detail report the experimental results in real-world dataset and applications.

\subsection{Experimental Setttings}
\vspace{1mm}

\noindent \textbf{Datasets.} 
Extensive experiments are conducted in both the Unite States and Chinna's stock markets by choosing the constituted entities in S \& P 500 and CSI 300 index. 
The historical price data from 2016 to 2021 are chosen as our datasets.
In addition to historical price data, our input data also includes company relation graphs. The graphs are generated by the stock price correlation matrix, which is introduced in Section~\ref{sec:cormat}. 
The stock price correlation matrix of each day is determined by the historical price movement of past 20 trading days. 
Specifically, we compare the opening price, closing price, trading volume, and trading volume of each pair of two stocks, calculate the correlation coefficient between them and take the mean value as the element of the correlation matrix.

\begin{table*}[ht]
	\small
	\caption{Performance evaluation of compared models for financial time series prediction in S\&P 500 and CSI 300 datasets. ACC and ARR measure the prediction performance and portfolio return rate of each prediction model, respectively, where the higher is better. AV and MDD measure the investment risk of each prediction model where the lower absolute value is better. ASR, CR, and IR measure the profit under a unit of risk, where the higher is better.}\vspace{-10pt}
	\centering
    \tabcolsep 6pt
	\begin{center}
		\begin{tabular}{p{0.24\columnwidth}p{0.08\columnwidth}p{0.08\columnwidth}p{0.08\columnwidth}p{0.1\columnwidth}p{0.08\columnwidth}p{0.08\columnwidth}p{0.08\columnwidth}p{0.08\columnwidth}p{0.08\columnwidth}p{0.08\columnwidth}p{0.1\columnwidth}p{0.08\columnwidth}p{0.08\columnwidth}p{0.08\columnwidth}p{0.08\columnwidth}}  \toprule
			\multirow{2}{*}{} & \multicolumn{7}{c}{S\&P 500} & \multicolumn{7}{c}{CSI 300}\\\cmidrule(lr){2-8}\cmidrule(lr){9-15}& ACC & ARR & AV & MDD & ASR & CR & IR & ACC & ARR & AV & MDD & ASR & CR & IR \\
			\toprule
			LSTM & 0.532 & 0.377 & 0.449 & -0.382 & 0.842 & 0.989 & 0.954 &  0.515 & 0.291 & \textbf{0.318} & -0.240 & 0.915 & 1.213 & 0.877\\ 
			GRU & 0.530 & 0.362 & \textbf{0.445} & -0.379 & 0.813 & 0.955 & 0.934 &  0.517  & 0.312 & 0.320 & -0.243 & 0.975 & 1.284 & 0.932\\
			Transformer & 0.533 & 0.385 & 0.454 & -0.384 & 0.848 & 1.005 & 0.960 &  0.518 & 0.327 & 0.322 & -0.245 & 1.016 & 1.335 & 0.969\\ 
			eLSTM & 0.534 & 0.434 & 0.454 & -0.373 & 0.955 & 1.163 & 1.041 & 0.520 & 0.330 & 0.323 & -0.239 & 1.022 & 1.381 & 0.991\\ \midrule
			LSTM+GCN & 0.538 & 0.470 & 0.442 & -0.354 & 1.062 & 1.326 & 1.103 & 0.523 & 0.351 & 0.320 & \textbf{-0.217} & 1.097 & 1.618 & 1.119\\
			LSTM+RGCN & 0.565 & 0.558 & 0.463 & -0.366 & 1.205 & 1.522 & 1.203 & 0.536 & 0.509 & 0.326 & -0.235 & 1.561 & 2.166 & 1.537\\
			TGC & 0.552 & 0.528 & 0.455 & \textbf{-0.344} & 1.163 & 1.535 & 1.180 & 0.531 & 0.453 & 0.323 & -0.224 & 1.402 & 2.022 & 1.412\\
			MAN-SF & 0.551 & 0.527 & 0.467 & -0.357 & 1.130 & 1.478 & 1.157 & 0.527 & 0.418 & 0.334 & -0.225 & 1.251 & 1.858 & 1.282\\
			HATS & 0.541 & 0.494 & 0.466 & -0.387 & 1.060 & 1.277 & 1.110 & 0.525 & 0.385 & 0.332 & -0.249 & 1.160 & 1.546 & 1.116\\
			REST & 0.549 & 0.502 & 0.466 & -0.359 & 1.079 & 1.398 & 1.117 & 0.528 & 0.425 & 0.331 & -0.228 & 1.284 & 1.864 & 1.298\\
			AD-GAT & 0.564 & 0.535 & 0.457 & -0.371 & 1.170 & 1.444 & 1.187 & 0.539 & 0.537 & 0.329 & -0.240 & 1.632 & 2.238 & 1.596\\\midrule
			THGNN & \textbf{0.579} & \textbf{0.665} & 0.468 & -0.369 & \textbf{1.421} & \textbf{1.804} & \textbf{1.340} & \textbf{0.551} & \textbf{0.632} & 0.336 & -0.237 & \textbf{1.881} & \textbf{2.667} & \textbf{1.875}\\
			
			\bottomrule
		\end{tabular}
	\end{center}
	\label{tb:performance}\vspace{-10pt}
\end{table*}

\vspace{1mm}
\noindent \textbf{Parameter Settings.} 
Our temporal graph $\Tilde{G}$ contains company relationships for 20 trading days. The $d_{\mathcal{N}}$ and $t_{\mathcal{N}}$ of neighborhood function $\mathcal{N}(\cdot)$ are both set as $1$. During the graph generation process, the threshold for generating one edge is set as $0.6$.
The historical price data of the previous 20 trading days are used as input features.
The feature dimension $d_{feat}$ of the encoding layer is 6, the input dimension $d_{in}$ and encoding dimension $d_{enc}$ of the encoding layer are both 128. The hidden dimension $d_{hidden}$ is 512, the dimension of value $d_v$ is 128, and the number of heads $h_{enc}$ is 8. 
In temporal graph attention layer, $d_{att}$ is 256, and the number of heads $h_{tga}$ is 4. 
The dimension of the attention vector in the heterogeneous graph attention layer, $d_q$, is 256.

\vspace{1mm}
\noindent \textbf{Trading Protocols.} 
\label{sec:protocol}
On the basis of~\cite{Dixon2017ClassificationbasedFM}, we use the daily buy-hold-sell trading strategy to evaluate the performance of stock movement prediction methods in terms of returns.
During each trading day during the test period (from January 1, 2020 to December 31, 2020), we use simulated stock traders to predict transactions:
\begin{enumerate}
    \item When the trading day $t$ closes, traders use this method to get the prediction score, that is, the ranking of the predicted rate of return of each stock.
    \item When the trading day $t+1$ opens: the trader sells the stock bought on the trading day $t$. Meanwhile, traders buy stocks with high expected returns, i.e., the stocks with top-$k$ scores.
    \item Please note that if a stock is continuously rated with the highest expected return, the trader holds the stock.
\end{enumerate}

In calculating the cumulative return on investment, we follow several simple assumptions:
\begin{enumerate}
    \item Traders spend the same amount on each trading day (for example, \$50,000). We made this assumption to eliminate the time dependence of the testing process in order to make a fair comparison.
    \item There is always enough liquidity in the market to satisfy the opening price of the order on the $t+1$ day, and the selling price is also the opening price on the $t+1$ day.
    \item Transaction costs are ignored because the cost of trading US stocks through brokers is relatively cheap, whether on a transaction-by-transaction basis or on a stock-by-stock basis.Fidelity Investments (Fidelity Investments) and Interactive Brokerage (Interactive Broker), for example, charge \$4.95 and \$0.005 per transaction, respectively.
\end{enumerate}

\vspace{1mm}
\noindent \textbf{Compared Baselines.} 
We compared our proposed method with state-of-the-art sequential-based models as well as the graph-based approaches. They are 1) Non-graph-based Methods, includes LSTM \cite{Hochreiter1997LongSM},  GRU~\cite{Cho2014LearningPR}, 
Transformer~\cite{Vaswani2017AttentionIA}, eLSTM~\cite{Li2020Elstm}. 2) Graph-based Methods: LSTM-GCN~\cite{Chen2018IncorporatingCR}, LSTM-RGCN~\cite{Li2020ModelingTS}, TGC~\cite{Feng2019TemporalRR}, MAN-SF~\cite{Sawhney2020DeepAL}, HATS~\cite{Kim2019HATSAH}, REST~\cite{Xu2021RESTRE} and  AD-GAT~\cite{Cheng2021ModelingTM}.

\vspace{1mm}

\noindent \textbf{Evaluating Metrics.} 
Since the goal is to accurately select the stocks with highest returns appropriately, we use seven metrics: prediction accuracy (ACC), annual return rate (ARR), annual volatility (AV), maximum draw down (MDD), annual sharpe ratio (ASR), calmar ratio (CR), and information ratio (IR) to report the performance of baselines and our proposed model.
Prediction accuracy is widely used to evaluate classification tasks, such as stock movement prediction~\cite{Li2020Elstm,Chen2019InvestmentBC}, so we calculate the prediction accuracy of all stocks for each trading day during the test period.
Because ARR directly reflects the effect of stock investment strategy, ARR is our main measure, which is calculated by adding up the returns of selected stocks on each test day in a year.
AV directly reflects the average risk of investment strategy per unit time. MDD reflects the maximal draw down of investment strategy during the whole test period. ASR reflects the benefits of taking on a unit of volatility with $ASR=\frac{ARR}{AV}$. CR reflects the benefits of taking on a unit of draw down with $CR=\frac{ARR}{\text{abs}(MDD)}$. IR reflects the excess return under additional risk.
The smaller the absolute value of AV and MDD is, the higher the value of ACC, ARR, ASR, CR, and IR is, the better the performance is.
For each method, we repeated the test process five times and reported average performance to eliminate fluctuations caused by different initialization.

\begin{figure*}[tb!]\vspace{-10pt}
   \centering
   \includegraphics[height=120pt,width=0.99\linewidth]{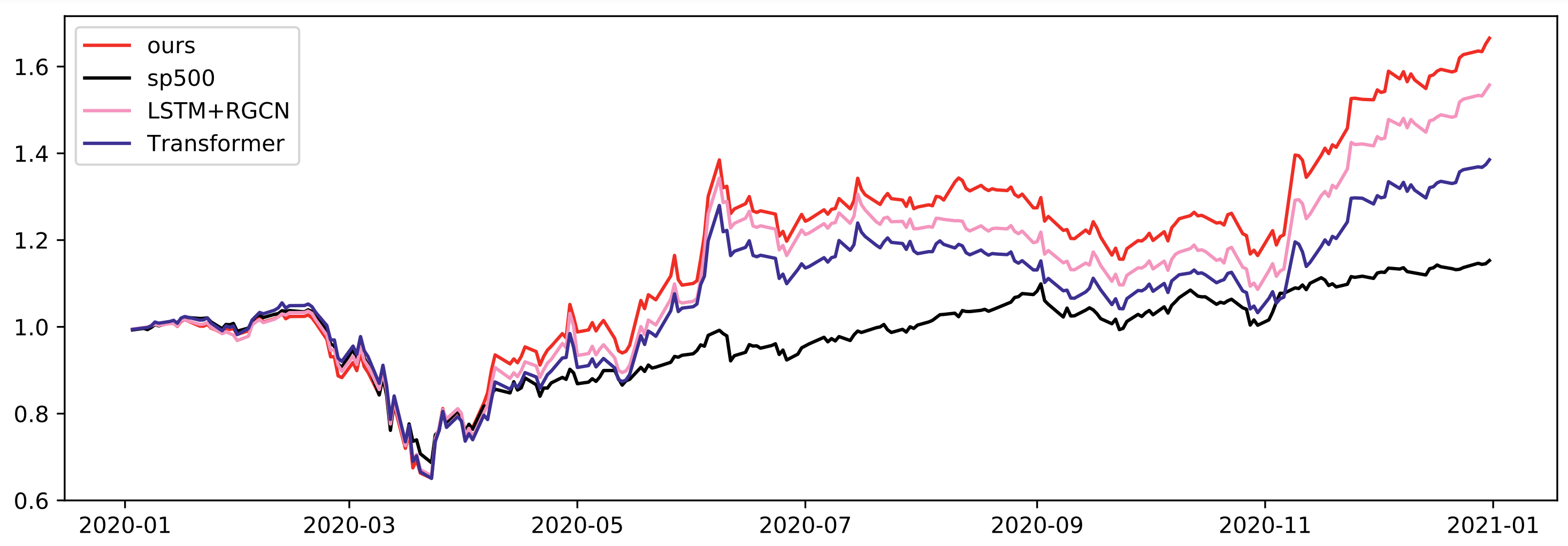}\vspace{-10pt}
   \caption{
   The accumulated returns gained in the test set (2020) by our proposed THGNN and selected baselines. For better illustration, we select one baseline from non-graph-based model and graph-based model, respectively.
   }
   \label{fig:portfolio}\vspace{-10pt}
\end{figure*}

\subsection{Financial Prediction}
In this section, we evaluate the performance of financial time series prediction and portfolio, which is the main task of this paper. 
Table~\ref{tb:performance} reports the performance through evaluation metrics such as ACC and ARR for each method in two datasets. 
The first four rows of Table~\ref{tb:performance} show the performance of models that do not use graph-based technology. 
It it clear that, none of these four methods is satisfactory, and their performance is all lower than that of other baselines. 
This proves that models without using company relationship data cannot achieve optimal performance. 

Lines 5 to 11 of Table~\ref{tb:performance} show the performance of the baseline models using graph-based technology. 
According to line 6, LSTM+RGCN performs best. 
This proves the effectiveness of using heterogeneous graphs of inter-company relationships. 
Note that according to line 7, TGC's performance is also competitive and its investment strategy is less volatile. 
This proves the effectiveness of using the dynamic relationship between companies.

According to the previous observation, the financial prediction model can be improved by using the heterogeneous or dynamic relationships of the company relation graph. Therefore, it is necessary to design an innovative model to improve the prediction performance of financial series from both dynamic and heterogeneous graph structure. 
According to the last row of Table~\ref{tb:performance}, our proposed THGNN outperforms all baselines and proves the superiority of temporal and heterogeneous graph neural network in financial time series prediction.

\begin{table}\vspace{-0pt}
	\small
	\caption{Performance evaluation of ablated models for financial time series prediction in S\&P 500 dataset. ACC, ARR, ASR, CR, and IR measure the prediction performance and portfolio return rate of each prediction model, where the higher is better.}\vspace{-10pt}
	\centering\small
    \tabcolsep 3pt
	\begin{center}
    \resizebox{0.45\textwidth}{!}{
		\begin{tabular}{p{0.32\columnwidth}p{0.08\columnwidth}p{0.08\columnwidth}p{0.08\columnwidth}p{0.1\columnwidth}p{0.08\columnwidth}}  \toprule
			& ACC & ARR & ASR & CR & IR \\
			\toprule
			THGNN-noenc & 0.548 & 0.571 & 1.201 & 1.524 & 1.082 \\ 
			THGNN-notemp & 0.539 & 0.486 & 0.964 & 1.292 & 0.946 \\ 
			THGNN-nohete & 0.553 & 0.600 & 1.279 & 1.618 & 1.198 \\ \midrule
			THGNN & \textbf{0.579} & \textbf{0.665} & \textbf{1.421} & \textbf{1.804} & \textbf{1.340} \\
			
			\bottomrule
		\end{tabular}}
	\end{center}
	\label{tb:ablation}\vspace{-10pt}
\end{table}

\subsection{Ablation Study}
In this section, we conduct the ablation experiments, that is, evaluating the performance of our methods that one part is dropped. 

According to the first row of Table~\ref{tb:ablation}, THGNN-noenc can not achieve the best performance after dropping the historical price encoding layer. 
This is because the encoder is responsible for extracting the time correlation in the historical price information. 
According to the second row, THGNN-notemp achieves unsatisfactory performance after dropping the temporal graph attention layer. 
This is because the temporal graph attention layer is responsible for dynamically adjusting the relationship between companies. 
Moreover, the relationship between companies changes dynamically over time, especially over a long period of time. 
According to the third row, THGNN-nohete cannot achieve the best performance after dropping the heterogeneous attention mechanism. 
This is because the heterogeneous graph attention layer is responsible for weighing the importance of messages from different relationships.

\subsection{Performance of the Portfolio}

In the performance evaluation of the portfolio strategy, we reported 6 widely-used evaluating metrics for portfolio, e.g., the annualized rate of return (ARR), annual sharp ratio (ASR), calmar ratio (CR), and information ratio (IR). 
Then, we show the accumulative return curve to compare the investment portfolio of our model and baselines during the test period. 
According to the trading protocols mentioned by section~\ref{sec:protocol}, we use the output of the prediction model to adjust our position day by day. 

Table~\ref{tb:performance} reports the performance of our model's and baselines' portfolio strategies. 
It is clear that our method has achieved the best performance in four of the six investment evaluating metrics. 
Specifically, our method performs best in terms of ARR, ASR, CR and IR. 
TGC and LSTM+GCN perform better than our model in terms of AV and MDD. 
This shows that our proposed THGNN takes the initiative to take more risks to pursue higher risk-return ratio than other baselines'.
According to Table~\ref{tb:ablation}, THGNN outperforms its sub-models in terms of portfolio strategy return evaluating metrics (e.g., ARR, ASR, CR, and IR). 
Therefore, our model shows strong effectiveness in building profitable portfolio strategies.

In order to further establish and evaluate the returns of our model during the test time, we calculate the cumulative returns for each trading day during the test period. 
We report the cumulative return curve of our model and other models on each trading day in Figure~\ref{fig:portfolio}. 
Due to space constraints, we select some representative baselines to compare with our model. 
We can observe that all baselines outperform the S\&P500 index. 
None of the models beat the others in the first four months. 
After starting in May, our model began to stay ahead of other models. 
In the following time, our model gradually widened the gap with other models. 
THGNN remained in the lead in the last month of 2020 and eventually achieved a profit on investment of more than 60 per cent.
Experimental results on other baselines can draw similar conclusion.

\subsection{Parameter Sensitivity}
\begin{figure}[tb!]\vspace{-0pt}
\centering
\subfigure[Dimension of final embedding $d_{att}$]{
\begin{minipage}[t]{0.5\linewidth}
\centering
\includegraphics[width=1.7in]{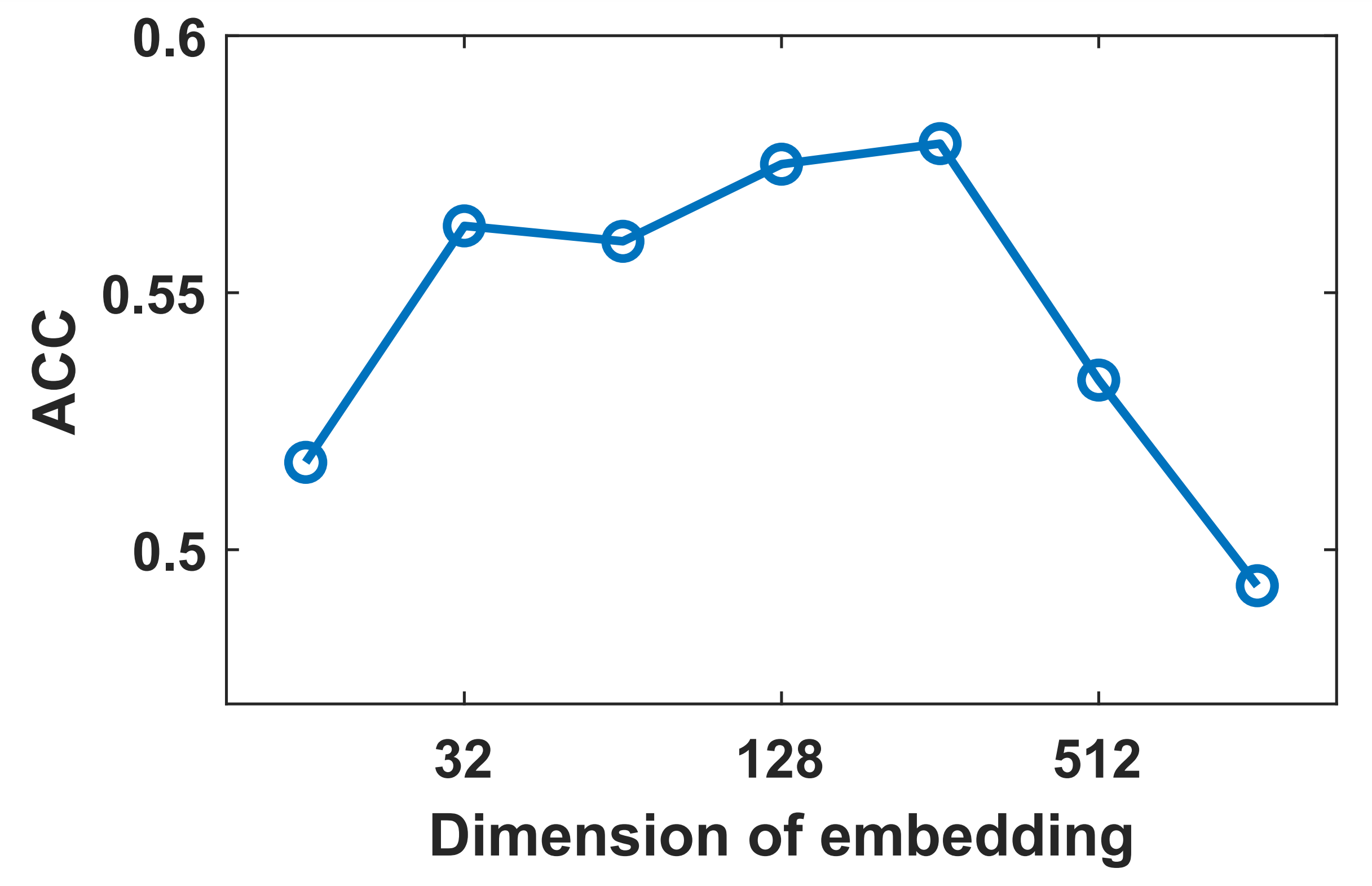}\vspace{-5pt}
\end{minipage}%
\vspace{-10pt}
}%
\subfigure[Dimension of encoding output $d_{enc}$]{
\begin{minipage}[t]{0.5\linewidth}
\centering
\includegraphics[width=1.7in]{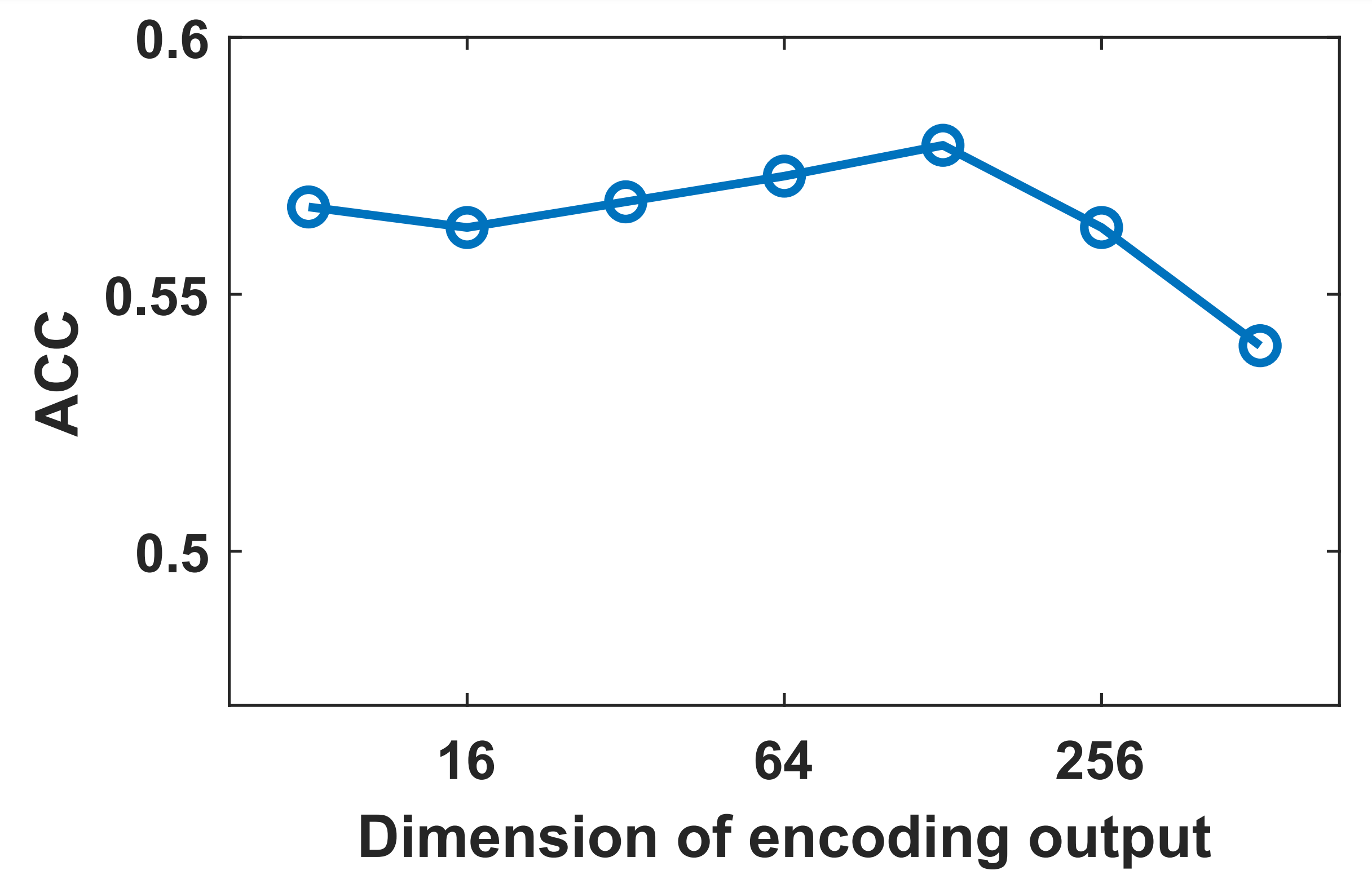}\vspace{-5pt}
\end{minipage}%
\vspace{-10pt}
}%

\subfigure[Dimension of the attention vector $\mathbf{q}$]{
\begin{minipage}[t]{0.5\linewidth}
\centering
\includegraphics[width=1.7in]{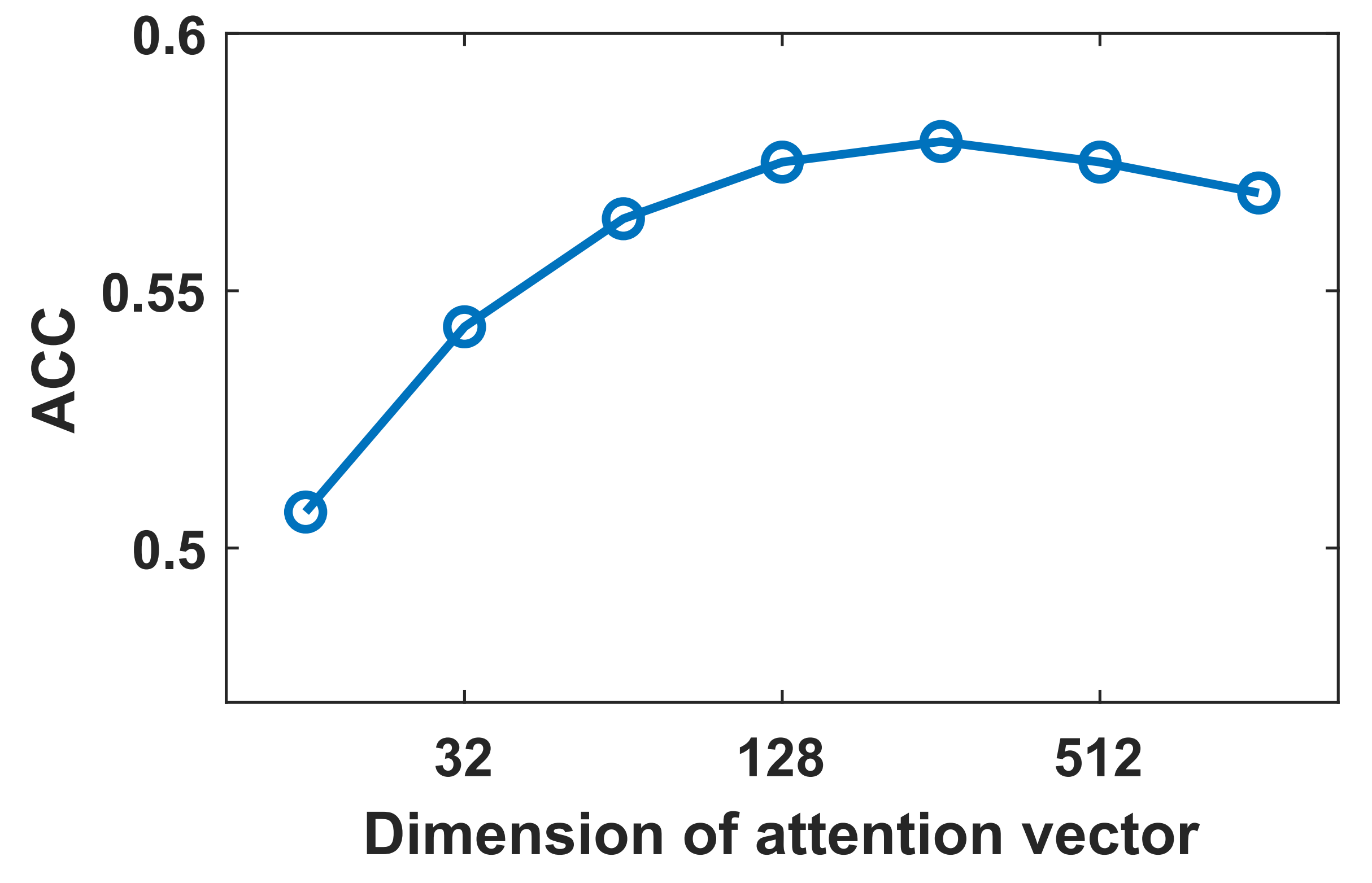}\vspace{-5pt}
\end{minipage}%
\vspace{-10pt}
}%
\subfigure[Number of attention head $h$]{
\begin{minipage}[t]{0.5\linewidth}
\centering
\includegraphics[width=1.7in]{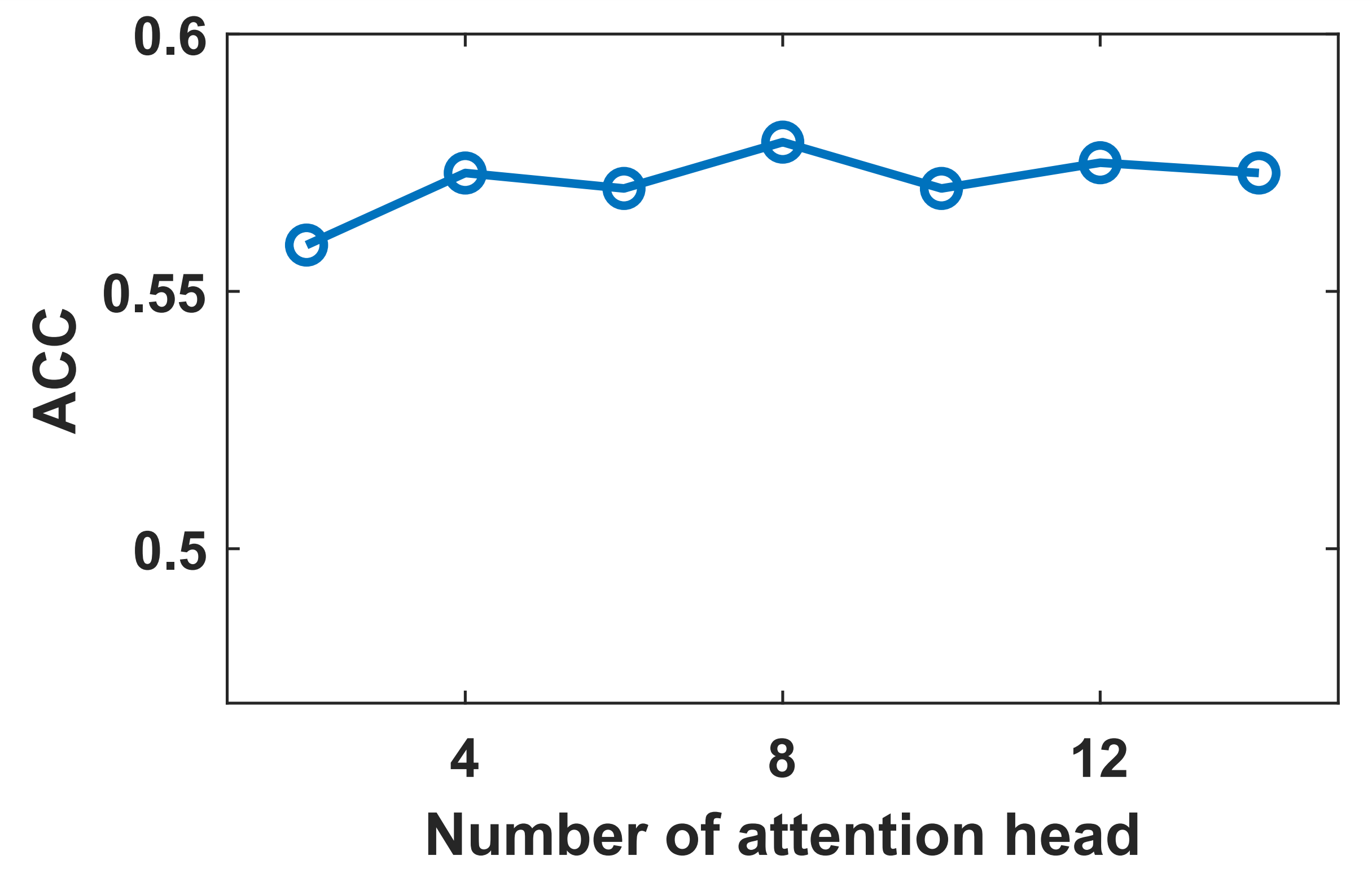}\vspace{-5pt}
\end{minipage}%
\vspace{-10pt}
}%
\caption{
Prediction performance of THGNN in terms of dimension of final embedding $d_{att}$, dimension of encoding output $d_{enc}$, dimension of the attention vector $\mathbf{q}$, and number of attention head $h$.
}\vspace{-10pt}
\label{fig:param}
\centering
\end{figure}

In this section, we report the experimental results of parameter sensitivity on the financial prediction task on S\&P500 dataset with various parameters in Figure~\ref{fig:param}.

According to Figure~\ref{fig:param} (a), we can observe that the performance of the model increases with the increase of embedded dimensions. 
The performance of the model peaked at 256 and then degraded rapidly. 
This is because embedded information needs a suitable dimension to reduce information loss and noise. 
According to Figure~\ref{fig:param} (b), the performance of the model slowly improves as the coding output dimension increases. 
The performance of the model begins to deteriorate after the dimension reaches 128. 
This is because the input information dimension is low, so the low-dimensional output can also achieve better performance. 
According to Figure~\ref{fig:param} (c), the performance of the model increases with the increase of the attention vector dimension. 
When the dimension reaches 256, the performance of the model reaches its peak. 
Continuing to increase the dimension will lead to overfitting, which will degrade the model. 
According to Figure~\ref{fig:param} (d), we can observe that the fluctuation of the model performance is low. 
We choose to pay attention to the number of force heads as 8. 
We also note that increasing the number of attention heads will make the model training more stable.

\subsection{Interpretability of Graph Neural Network}
The stock price fluctuation of listed companies is not only related to the historical price information of the company, but also affected by the companies related to it. 
Therefore, we need to integrate dynamic and heterogeneous relevant information as input to the prediction model. 
In our technology, the relationship between each company changes dynamically over time. 
The strength of the relationship between companies, that is, the proportion of contribution to the message also changes over time. 
Our time chart attention mechanism can dynamically adjust the importance of each company in the diagram. 
In addition, our heterogeneous attention mechanism can dynamically adjust the importance of each source. 
Therefore, our model can help to predict stock price volatility more accurately, and the experimental results verify the superiority of our model performance.

\begin{figure}[tb!]\vspace{-0pt}
\centering
\subfigure[Graph Attention in $pos$ Relationship]{
\begin{minipage}[t]{1\linewidth}
\centering
\includegraphics[height=0.8in,width=3.6in]{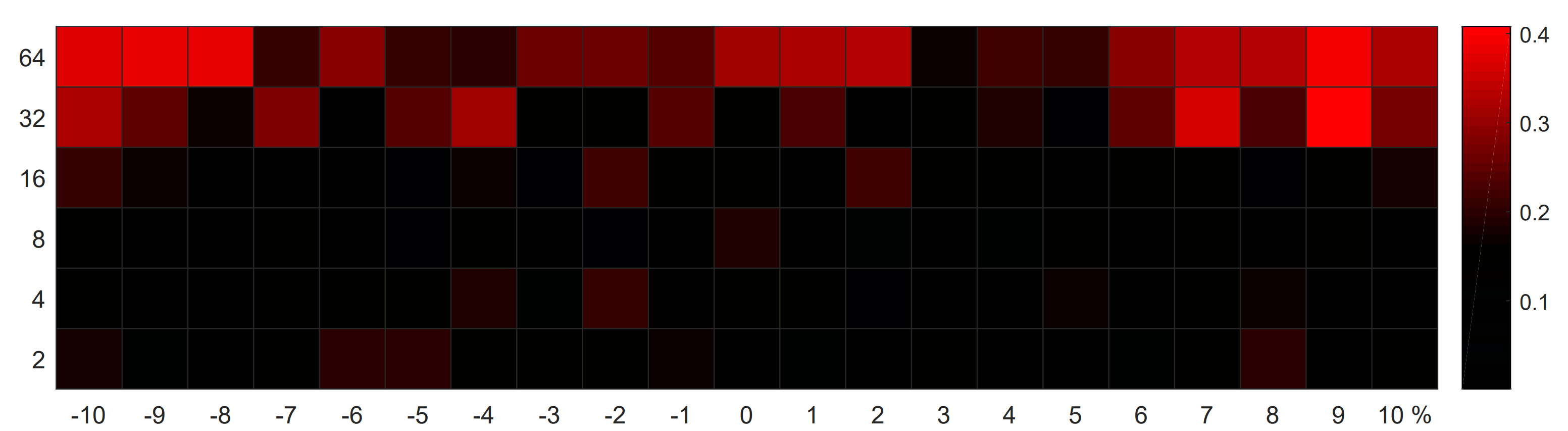}
\end{minipage}%
}%
\vspace{-10pt}
\subfigure[Graph Attention in $neg$ Relationship]{
\begin{minipage}[t]{1\linewidth}
\centering
\includegraphics[height=0.8in,width=3.6in]{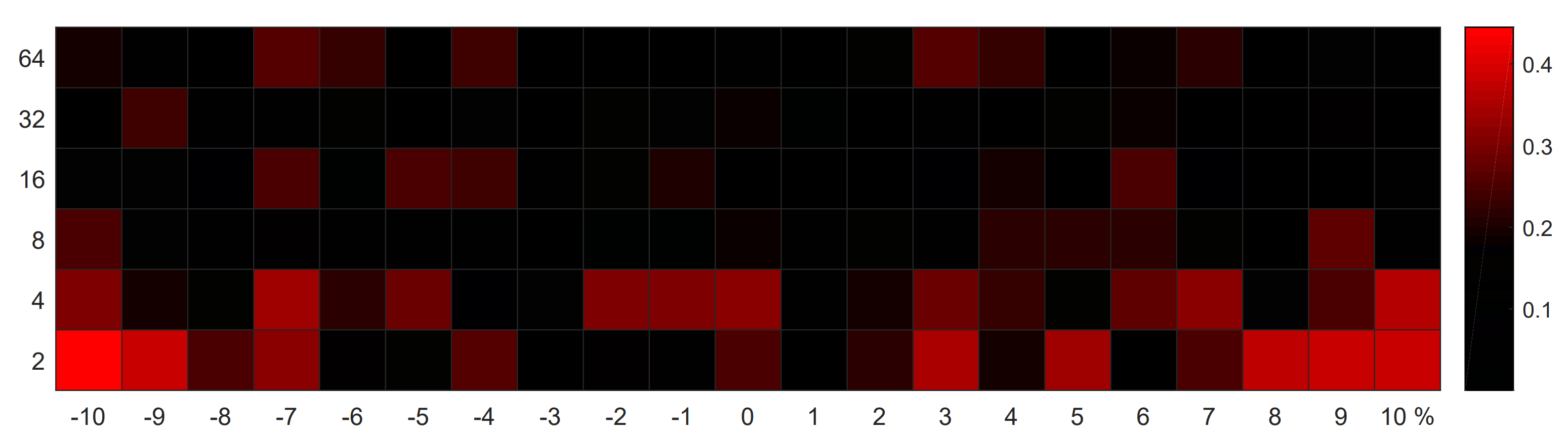}
\end{minipage}%
}%
\vspace{-10pt}
\caption{
Visualization of attention weights, X-axis denotes the daily return of stocks. Y-axis denotes the average degree in each company relation graph.
}\vspace{-10pt}
\label{fig:gat}
\centering
\end{figure}

Then, in order to explore the interpretability of our proposed model, we extract the attention weight of the graph in the process of the model prediction. 
We counted the attention weight of all nodes in the process of message delivery on the relational graph. 
Under different daily returns and node degrees, we take the mean value of attention weights and visualize the statistical results, which are reported in Figure~\ref{fig:gat}.

\begin{figure}[tb!]\vspace{-5pt}
\centering
\subfigure[ACC values on S\&P500.]{
\begin{minipage}[t]{0.5\linewidth}
\centering
\includegraphics[width=1.7in]{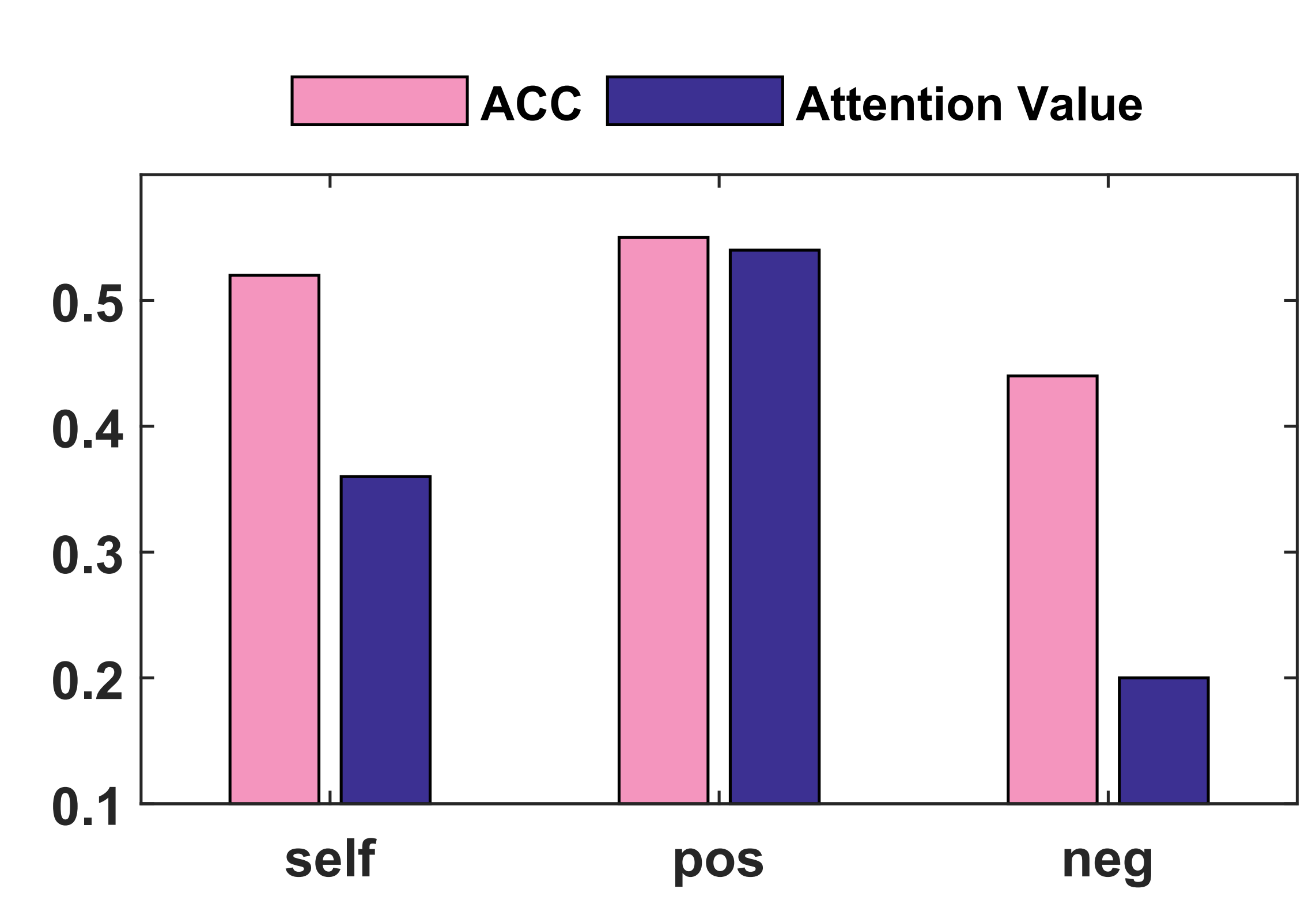}
\end{minipage}%
}%
\subfigure[ACC values on CSI300.]{
\begin{minipage}[t]{0.5\linewidth}
\centering
\includegraphics[width=1.7in]{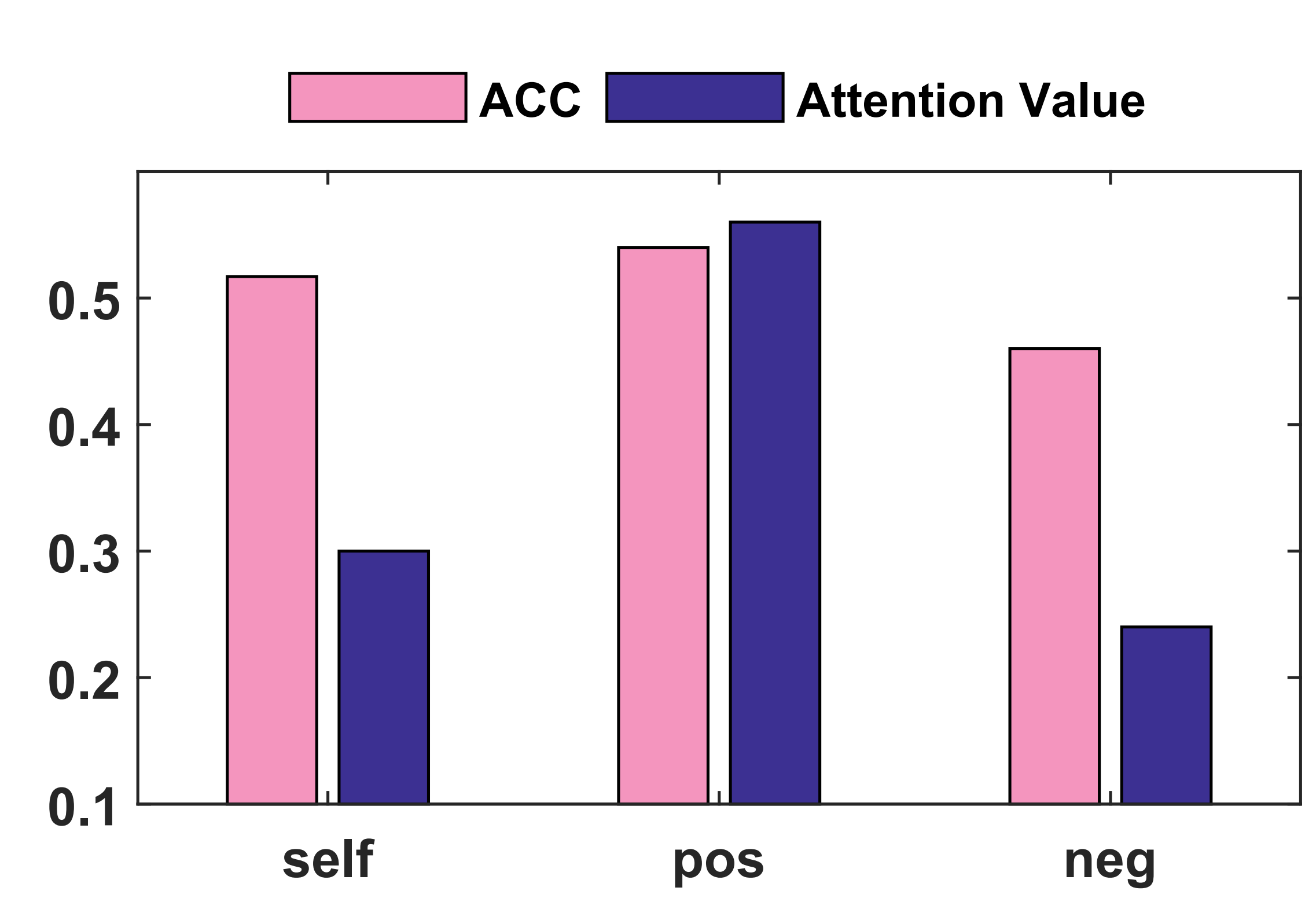}
\end{minipage}%
}%
\caption{
Prediction performance of single message source and corresponding attention value.
}\vspace{-10pt}
\label{fig:hete}
\centering
\end{figure}

\begin{figure*}[tb!]\vspace{-10pt}
   \centering
   \includegraphics[width=0.98\linewidth]{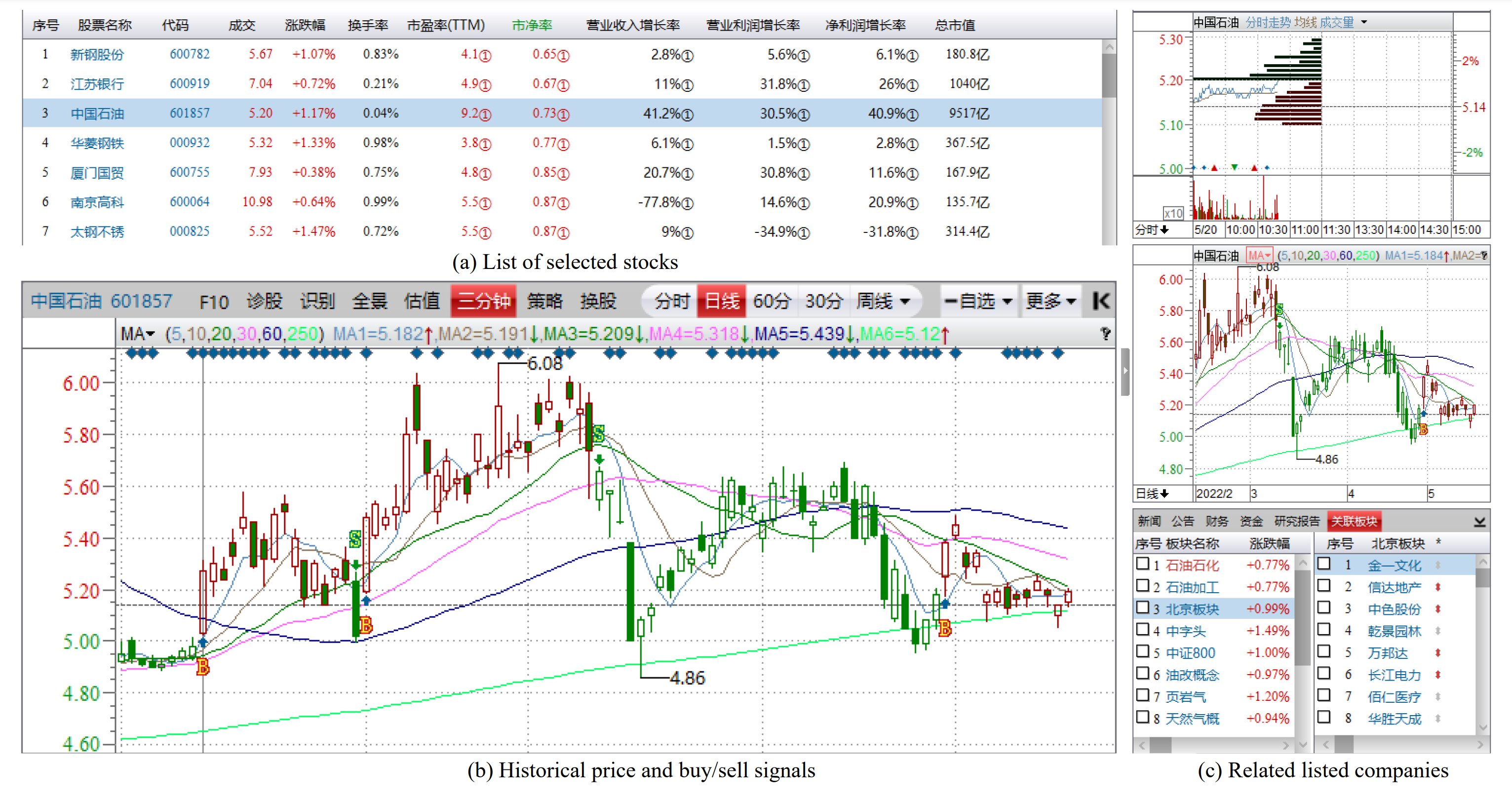}\vspace{-10pt}
   \caption{
   The desktop interface of investment portfolio based on our proposed THGNN method. It includes price-relevant listed companies from historical data and visualization of how does our method makes predictions on buying and selling. The part (a) lists the stocks held by our method in order from highest to lowest. And part (b) shows the 'buy' and 'sell' signals generated by our trading protocols. Then part (c) lists the listed companies related to China National Petroleum Corporation (CBPC: 601857) and shows which ones have higher correlation to this company according to our generated stock correlation graph.
   }
   \label{fig:strategy}\vspace{-0pt}
\end{figure*}

Figure~\ref{fig:gat} (a) shows the attention weights on the $pos$ diagram. 
We can see (on the y-axis) that the nodes with higher degrees have higher average attention weights. 
This shows that in the process of message delivery on the $pos$ graph, the degree is higher, that is, the nodes with more neighbors will contribute more messages to their neighbors. 
We also found that (on the x-axis), companies with larger fluctuations in daily returns also had higher average attention weights. 
This shows that more volatile price changes will contribute more information to their neighbors, which also means that price fluctuations will produce momentum spillover effects. 
According to Figure~\ref{fig:gat} (b), we can see that on the $neg$ graph, with a lower degree node, the average attention weight is higher. 
This indicates that during message delivery on the $neg$ graph, nodes with lower degrees contribute more messages to their neighbors.

For more interpretable experimental results, we also visualize each relationship's attention weights and show the corresponding performance when using only one relationship. Specifically, we trained our proposed THGNN at two datasets, and counted the mean value of the attention weights in heterogeneous graph attention layer. Then, we used each single relationship's message as the input of the prediction model to obtain the prediction performance, as illustrated in Figure~\ref{fig:hete}.

It is clear that $self$ and $pos$ message resources achieve better prediction performance than $neg$. Moreover, the $pos$ message resource contribute importantly to the prediction model, which proves the contribution to the prediction model. The reason might be that the influence between companies in the similar price movement is relatively useful for prediction future price movement. Although the $neg$ message resource has an unsatisfactory performance when predicting price movement single, it still contributes to our proposed THGNN in achieving the state-of-the-art performance according to Table~\ref{tb:ablation}. 
We can see that temporal graph attention layer can reveal the difference between nodes and weights then adequately, and the heterogeneous graph attention can adjust the contribution of each message resource adaptively. The result demonstrates the effectiveness of graph-structure information and the interpretability of proposed graph neural network model.

\subsection{System Implementation}
In this section, we introduce the implementation detail of our proposed methods. We first show the settings of model employment and training strategy. Then we show the web-based application, which shows how our proposed method gives customers informative advice. Our proposed THGNN is re-trained every trading day. To handle the large-scale dataset, we leverage mini-batch gradient descent optimization with a batch size of 256 and a learning rate of 3e-4. The model is implemented by PyTorch and deployed in Python and Java. Besides, we use distributed Scrapy~\cite{Myers2015ChoosingS} to obtain historical stock data and utilize Neo4j~\cite{Holzschuher2013PerformanceOG} as the graph database to store relational graphs.

Figure~\ref{fig:strategy} shows the interface of our desktop application. 
The upper left part of Figure~\ref{fig:strategy} is the list of stocks to be held based on the THGNN strategy.
The lower left part of Figure~\ref{fig:strategy} reports the price change curve of China National Petroleum Corporation (CBPC: 601857). It contains buy and sell points suggested by our THGNN. B denotes buying and S denotes selling. It can be seen that our model provides three buy signals and two sell signals.
The lower right part of Figure~\ref{fig:strategy} reports the relevant companies of CBPC. Companies with high stock price volatility correlations are marked with red arrowhead. 
The results shows that our investment strategy provides informative advice through a relational graph approach.

\section{Conclusion and Discussion}
In this paper, a new temporal and heterogeneous graph neural network model is proposed for financial time series prediction. 
Our method addresses the limitations of the existing works of graph neural networks by adjusting the message contribution ratio of each node through temporal and heterogeneous graph attention mechanism. 
We evaluate the effectiveness of the proposed method comprehensively by comparing it with the most influential graph-based and non-graph-based baselines. 
In addition, THGNN performs better than other baselines in the actual investment strategy, and the results show that our approach based on dynamic heterogeneous graph  can obtain a more profitable portfolio strategy than those based on static or homogeneous graph structure. 

In conclusion, this paper is the first time to model the inter-company relationship into a heterogeneous dynamic graph, and apply it to the financial time series prediction problem. 
This is beneficial to the more extensive research and innovation of graph-based technology in the financial field. 
On the one hand, we model the company relation graph as the real dynamic heterogeneous graph; on the other hand, we improve the financial time series prediction model through the latest graph neural network technology. Besides, there is still room for improvement in our work on generating real-life company relation graphs.
In the future, we will focus on improving the modeling of the company's relation to help the prediction model obtain more accurate training input graph data.

\section*{Acknowledgments}
Dawei Cheng is supported by the NSFC 62102287, Ying Zhang is supported by ARC DP210101393.

\bibliographystyle{ACM-Reference-Format}
\balance
\bibliography{acmart}

\end{document}